\documentclass[preprint,aps,superscriptaddress,nofootinbib]{revtex4}

\usepackage{graphicx}

\def\slash#1{#1\!\!\!/\!\,\,}
\def\nb{\bar n}
\def\bn{\bar n}
\def\pslash{p\:\!\slash{}\!}
\def\ppslash{\pslash'\!\!{}}
\newcommand{\nslash}{\slash n}

\newcommand{\partialslash}{\slash \partial\!}

\newcommand{\bnP}{\bar {\cal P}}

\newcommand{\cPslash}{{\cal P}\:\!\!\slash{}\!}
\newcommand{\Dslash}{D\!\slash{}}
\newcommand{\LQCD}{\Lambda_{\rm QCD}}
\newcommand{\SCETa}{SCET$_{\rm I}$}
\newcommand{\SCETb}{SCET$_{\rm II}$}

\begin{document}

\preprint{\vbox{\hbox{FERMILAB-Pub-03/035-T} \hbox{LBNL-52196} 
\hbox{CALT-68-2432} \hbox{hep-ph/0303099} }}

\title{Comment on Quark Masses in SCET}

\author{Adam K.\ Leibovich}\email{adam@fnal.gov}
\affiliation{Department of Physics and Astronomy, University of Pittsburgh,
        Pittsburgh PA 15260\vspace{6pt} }
\affiliation{Theory Group, Fermi National Accelerator Laboratory,
        Batavia IL 60510\vspace{6pt} }
	
\author{Zoltan Ligeti}\email{zligeti@lbl.gov}
\affiliation{Ernest Orlando Lawrence Berkeley National Laboratory,
        University of California, Berkeley CA 94720\vspace{6pt} }

\author{Mark B.\ Wise\,}\email{wise@theory.caltech.edu}
\affiliation{California Institute of Technology, Pasadena, CA 91125
        \\ $\phantom{}$ }

\vspace*{1.5cm}

\begin{abstract}

Quark masses are included in the SCET Lagrangian.  Treating the strange quark
mass as order $\LQCD$, we find that strange quark mass terms are suppressed in
\SCETa, but are leading order in \SCETb.  This is relevant for $B$ decays to
$K^*$ and $K$.  Strange quark mass effects in semileptonic and weak radiative
form factors are studied. They give corrections to the form factors that are
not suppressed by powers of the bottom quark mass, or, equivalently, by the
large recoil energy of the final state meson, and preserve the heavy to light form factor relations that follow from using the leading order current.

\end{abstract}

\maketitle

\newpage

Our understanding of processes involving a collinear  jet of particles has
improved significantly recently  due to the construction of the Soft-Collinear
Effective Theory (SCET)~\cite{SCET,Bauer:2000yr,Bauer:2001ct,factorization}. 
The effective field theory couples soft physics to highly energetic quarks and
gluons moving in a collinear jet.  The symmetries of the SCET simplify proofs
of factorization~\cite{Bauer:2001ct,factorization,Bauer:2002nz} and
calculations of Sudakov logarithms.  SCET has been applied to many processes
including $B$ meson \cite{SCET, Bauer:2000yr, factorization,Beneke,
otherBpapers} and quarkonium \cite{quarkonium} decays, jet physics \cite{jet},
and the pion form factor \cite{Rothstein:2003wh}.  Light quark mass terms in
SCET were first considered in Ref.~\cite{Rothstein:2003wh}, and our work
elaborates on the discussion there.

The lightcone components of the collinear particles scale as $p =
(p^+,p^-,p_\perp) \approx Q(\lambda^2,1,\lambda)$, where $Q$ is the large
energy scale, and the expansion parameter $\lambda\ll 1$ depends on the
particular process.  For example, $\lambda = \sqrt{1-2E_\gamma/M}$ for
inclusive meson decays (i.e., $B$ or $\Upsilon$) to a photon.  In $B$ decays,
there are two appropriate choices, $\lambda = \sqrt{\LQCD/m_b}$ or $\lambda =
\LQCD/m_b$.  To distinguish between the two cases, the effective theories are
called \SCETa\ and \SCETb, respectively~\cite{Bauer:2002aj}.

Factorization is proven in SCET by using a field redefinition which decouples
soft quanta from collinear particles.  For example, in the proof of
factorization for $B\to D\pi$, after the soft physics is decoupled, there is a
matrix element of a pair of collinear quark fields which becomes the pion decay
constant~\cite{factorization}.  At leading order in $\lambda$, no soft physics
couples to the pion in the rest frame of the $B$ meson, and so the
nonperturbative strong interaction  physics that gives rise to confinement and
chiral symmetry breaking must exist in the collinear sector of the theory if
this approach to factorization is correct.

Intuitively, the light quark mass should be suppressed in the collinear
Lagrangian, since at very large energies, the quark behaves as if it were
massless.  However this leads to a problem.  Apart from a  factor of the
Cabibbo angle, the only difference between $\bar B^0\to D^+\pi^-$ and $\bar
B^0\to D^+K^-$ is the flavor of one of the final state collinear quarks.  If
light quark mass terms are suppressed  in the collinear Lagrangian then
differences in the flavor of light quarks can only enter in the soft physics,
which at leading order in $\lambda$ does not couple to the final state meson. 
However, the difference between the $\pi$ and $K$ decay constants is not small
(i.e., around 30\%), and furthermore this difference enters the  decay rates in
a way that is not  suppressed by  the available energy $1/(m_B-m_D)$.  So if
$m_s$ is treated as order $\LQCD$, the strange quark mass terms must be order
one in the collinear Lagrangian.

Apart from isospin violation, the up and down quark masses will always be
negligible, however the strange quark mass, $m_s\sim\LQCD$, could be
important.  Below we investigate how quark masses enter into SCET and under what
circumstances the strange quark mass is important.

Begin by defining two lightcone vectors, $n^\mu$ and $\nb^\mu$, such that $n^2
= \nb^2 = 0$ and $n\cdot\nb = 2$.  In SCET there are fundamental fields and
Wilson lines, which are built out of the fields.  Furthermore, there are two
separate sectors to the theory: collinear and ultrasoft (usoft).  In the
collinear sector there is a collinear fermion field $\xi_{n,p}$, which is
obtained by decomposing the full QCD quark field as
\begin{equation}\label{fieldredef}
\psi(x) = \sum_{\tilde p} e^{-i\tilde p\cdot x}
  \left(\frac{\slash n\slash\nb}{4} + \frac{\slash\nb\slash n}{4}\right)
  \psi_{n,p} 
\equiv \sum_{\tilde p} e^{-i\tilde p\cdot x}(\xi_{n,p} +
  \xi_{\nb,p})\,,
\end{equation}
a collinear gluon field $A_{n,q}^\mu$, and a collinear Wilson line
\begin{equation}
W_n(x) = \bigg[ \sum_{\rm perms} {\rm exp}
  \left( -g\, \frac{1}{\bnP}\, \nb \cdot A_{n,p}(x) \right) \bigg] \,.
\end{equation}
The subscripts on the collinear fields are the lightcone direction $n^\mu$, and
the large components of the lightcone momenta, $\tilde p = n\, (\nb\cdot p)/2 +
p_\perp$.   Derivatives acting on collinear fields are order $\lambda^2$, since
the $p^-$ and $p_\perp$ components have been removed.  The operator ${\cal
P}^\mu$ projects out the momentum label~\cite{Bauer:2001ct}.  For example
$\nb\cdot {\cal P}\, \xi_{n,p} \equiv \bnP \xi_{n,p} = \nb \cdot p\,
\xi_{n,p}$.  Functions of the operator $\bnP$ have the property
\begin{equation}
f(\bnP + i\nb\cdot D) = W_n f(\bnP) W_n^\dagger\,.
\end{equation}
Likewise in the usoft sector there is a usoft fermion field $q_s$, a usoft
gluon field $A^\mu_s$, and a usoft Wilson line $Y$. 

Operators in SCET are constructed from these objects such that they are gauge
invariant.  For example, under collinear-gauge transformations $\xi_{n,p} \to
U_n\, \xi_{n,p}$ and $W_n \to U_n\, W_n$, so
\begin{equation}
\chi_n \equiv W^\dagger_n\, \xi_{n,p}
\end{equation}
is collinear-gauge invariant. This combination, $\chi_n$, however, still
transforms under a usoft-gauge transformation $\chi_n\to V(x) \chi_n$.

The SCET collinear Lagrangian is obtained by substituting
Eq.~(\ref{fieldredef}) into the QCD Lagrangian, ${\cal L} = \bar\psi(i\Dslash -
m)\psi$, which gives
\begin{eqnarray}
{\cal L} &=& \bar\xi_{n,p'}\, in\cdot D\, \frac{\slash\nb}{2}\,\xi_{n,p} 
  + \bar\xi_{\nb,p'}\, (\cPslash_\perp + i\Dslash_\perp - m)\, \xi_{n,p}
\nonumber\\*
&+& \bar\xi_{n,p'}\, (\cPslash_\perp + i\Dslash_\perp - m)\, \xi_{\nb,p}
  + \bar\xi_{\nb,p'}\, (\bnP + i\nb\cdot D)\, \frac{\nslash}2\, \xi_{\nb,p}\,,
\end{eqnarray}
where we have used the convention where momentum labels are implicitly summed
over~\cite{Bauer:2001ct}.

\begin{figure}[t]
\begin{tabular}{ccl}
$(\tilde p,k)~$ \\[-12pt]
\includegraphics*[width=.2\textwidth]{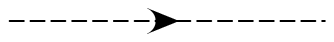}  & ~~~~ &  
  \raisebox{5pt}{$\displaystyle = i\, \frac{\nslash}{2}\,
  {\bn\cdot p \over n\cdot k\; \bn\cdot p + p_\perp^2 - m^2 +i\epsilon}$}
  \\[20pt]
\includegraphics*[width=.2\textwidth]{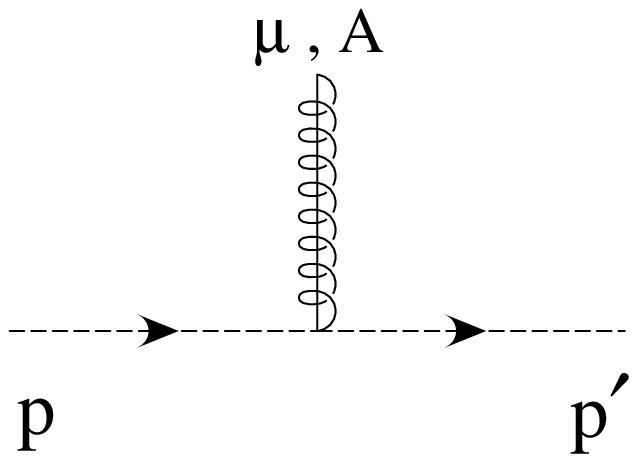} & &
  \raisebox{18pt}{\vbox{\hbox{$\displaystyle = i g\,T^A\,
    \bigg\{ n^\mu + \frac{ \gamma_\perp^\mu \pslash_\perp }{\nb \cdot p}
    + \frac{ \ppslash_\perp \gamma_\perp^\mu }{\nb \cdot p'}
    - \frac{\ppslash_\perp\, \pslash_\perp}{\bn \cdot p\;
    \bn \cdot p'}\, \bn^\mu $} \vspace*{5pt}
  \hbox{$\displaystyle\quad{} + \frac{m}{\bn\cdot p\; \bn\cdot p'}\Big[
    \gamma_\perp^\mu \big(\bn\cdot p' - \bn\cdot p\big) 
    + \bn^\mu(\pslash_\perp - \ppslash_\perp + m) \Big] 
    \bigg\}\, \frac{\bar\nslash}{2} $}} }
\end{tabular}
\caption{Order $\lambda^0$ Feynman rules: collinear quark propagator with label
$\tilde p$ and residual momentum $k$, and collinear quark interactions with one
collinear gluon, respectively.}
\label{feynmanrules}
\end{figure}

We can use the equations of motion to remove $\xi_{\nb,p}$,
\begin{equation}
(\bnP + i \nb\cdot D) \xi_{\nb,p} = 
  (\cPslash_\perp + i\Dslash_\perp + m)\frac{\slash\nb}{2}\, \xi_{n,p}\,,
\end{equation}
and make a field redefinition to remove the couplings to the (u)soft degrees of
freedom. This gives the usual leading order collinear Lagrange density
\begin{equation}\label{leadingL}
{\cal L}_0 = \bar\xi_{n,p'} \left\{in\cdot \partial + 
  (\cPslash_\perp + g \slash A_{n,q}^\perp) W_n \frac1\bnP W_n^\dagger
  (\cPslash_\perp + g \slash A_{n,q'}^\perp)\right\}
  \frac{\slash\nb}{2}\, \xi_{n,p} \,,
\end{equation}
and the following mass terms
\begin{equation}\label{masslagrangian}
{\cal L}_m = m\, \bar\xi_{n,p'}
  \left[(\cPslash_\perp + g \slash A_{n,q}^\perp),
  W_n\frac1\bnP W_n^\dagger\right] \frac{\slash\nb}{2}\,\xi_{n,p} 
- m^2\, \bar\xi_{n,p'} W_n \frac1\bnP W_n^\dagger\, 
  \frac{\slash\nb}{2}\,\xi_{n,p}\,.
\end{equation}
The Feynman rules for the collinear quark propagator and the interaction of a
collinear quark with a single  collinear gluon are show in
Fig.~\ref{feynmanrules}.

The next question is how $m$ scales.  If $m\sim Q\lambda^2$, the terms in
Eq.~(\ref{masslagrangian}) are suppressed and can be dropped from the collinear
Lagrangian at leading order.  Thus we would be left with only  ${\cal L}_0$,
which agrees with~\cite{Bauer:2000yr}.  The masses would still be important in
the usoft Lagrangian. If, however, $m\sim Q\lambda$, all terms in
Eq.~(\ref{masslagrangian}) are equally important compared with
Eq.~(\ref{leadingL}), and must be kept.  Note that when $m \sim Q\lambda \sim
\LQCD$, i.e., in \SCETb, the perpendicular label should be interpreted as a
derivative.  Thus Eq.~(\ref{masslagrangian}) becomes
\begin{equation}
\label{massterms}
{\cal L}_m = m\, \bar\xi_{n,p'}
  \left[(i\partialslash_\perp + g \slash A_{n,q}^\perp),
  W_n\frac1\bnP W_n^\dagger\right] \frac{\slash\nb}{2}\, \xi_{n,p} 
- m^2\, \bar\xi_{n,p'} W_n \frac1\bnP W_n^\dagger\, 
  \frac{\slash\nb}{2}\, \xi_{n,p}\,.
\end{equation}

The leading order collinear Lagrangian in Eq.~(\ref{leadingL}) has a $U(1)$
helicity symmetry which is generated by $\gamma_5$ acting on the collinear
quark field. However, this symmetry  is explicitly broken by the terms linear in
the quark mass in Eq.~(\ref{massterms}). It is also broken by nonperturbative
strong interaction physics which causes the vacuum expectation value,
\begin{equation}
\label{vev}
\langle \Omega|\,\bar\xi_{n,p'}
  \left[(i\partialslash_\perp + g \slash A_{n,q}^\perp),
  W_n\frac1\bnP W_n^\dagger\right] \frac{\slash\nb}{2}\,\xi_{n,p} 
  |\Omega \rangle\,,
\end{equation}
to differ from zero.
 
For inclusive $B \to X_s\gamma$ or $B\to X_u\ell\bar\nu$ decays, the
interesting region of phase space is $E \simeq m_b/2 - \LQCD$, giving $\lambda
= \sqrt{\LQCD/m_b}$\,, where we have taken $Q = {\cal O}(m_b)$.  The
appropriate effective theory is \SCETa, and therefore the dependence on the
light quark flavor is power suppressed in  these decays, implying the usual
relationship between the shape functions~\cite{shape}.

For $B$ decays to a light meson $M$, however, the invariant mass square of the
outgoing meson is of order $m_M^2 \sim m_b^2 \lambda^2 = {\cal O}(\LQCD^2)$,
where again we have taken $Q = {\cal O}(m_b)$.  Therefore, $\lambda =
\LQCD/m_b$ and \SCETb\ is the proper theory. Then, if $m_s \sim \LQCD$, we
must  include the strange quark mass effects in the leading order collinear
Lagrangian.  In practice it may be appropriate to treat the strange quark mass
as somewhat smaller than the strong interaction scale in which case it can be
treated as a perturbation. Then the term linear in $m_s$ in
Eq.~(\ref{massterms}) gives rise to corrections suppressed by $m_s/\LQCD$ but
not by powers of $\lambda$.  Our results show explicitly that $SU(3)$ breaking
in the relations between the $B\to K^*$ and $B\to \rho$ (or the $B\to K$ and
$B\to \pi$) form factors that describe semileptonic and weak radiative decays
is not suppressed by $\LQCD/m_b$ in any region of phase space.

In Ref.~\cite{Bauer:2002aj} the form factors for heavy to light weak
transitions are considered using SCET. They adopt a two step  process where one
first considers contributions in \SCETa\ and then matches onto \SCETb. The
terms that give a leading contribution to the form factors are suppressed by
$\lambda^2 = \LQCD/m_b$ in \SCETa. For $B_{u,d}$ decays, when the spectator
quark is not a strange quark, the terms in the mixed collinear-usoft Lagrangian
that cause the usoft spectator quark to transition to a collinear quark are the
same as in \cite{Bauer:2002aj}. These are suppressed by at least one power of
$\lambda$. As we have remarked earlier, in \SCETa\ the strange quark mass terms
are also suppressed by at least one power of $\lambda$.  Therefore, to get a
leading form factor contribution involving the strange quark mass term, the
time ordered products must contain the leading order current,
\begin{equation}
J^{(0)} = \bar \xi_n W_n \Gamma h_v,
\end{equation}
and these terms preserve the usual heavy to light form factor
relations~\cite{Charles}.  On transitioning to \SCETb, the strange quark mass
terms become leading order. However, the non-factorizable pieces which involve
$J^{(0)}$ will automatically preserve the form factor relations in \SCETb, even
with the strange quark mass term included in the leading order \SCETb\
Lagrangian.  In fact there is a simple physical argument that this should be
the case.  Constituent quark masses induced by chiral symmetry breaking often
act, in low energy phenomenology, much like explicit light quark masses.  Given
this it would be puzzling if explicit light quark masses violated the heavy to
light form factor relations that follow from only including the leading order
current, $J^{(0)}$, but spontaneous chiral symmetry breaking did not cause such
violations.  It is possible that the factorizable terms that violate the
leading order form factor relations are suppressed by a factor of
$\alpha_s(\sqrt{m_b \LQCD})$ compared to those that preserve
them~\cite{BF}.  If this is the case then there will be no
corrections to the form factor relations to any order in $m_s/\LQCD$. 

Finally we note that for the case of $B_s$ decays the strange quark mass terms
in \SCETa\ that cause the usoft spectator strange quark to transition to a
collinear one could be important.  However, using the equations of motion, it is not difficult to show that no terms of this type appear at order $\lambda$ or $\lambda^2$ \cite{Pirjol:2002km}.
Note that this does not mean that the differences between form factors for $B_s$ and $B$ decays are suppressed by powers of $\lambda$.

\acknowledgments

We are grateful to I.~Stewart, I.~Rothstein, and C.~Bauer for useful comments. 
A.K.L.\ and M.B.W.\ would like to thank the LBL theory group for its
hospitality while portions of this work were done.   A.K.L.~was supported in
part by the Department of Energy under Grant DE-AC02-76CH03000.  Z.L.~was
supported in part by the Director, Office of Science, Office of High Energy and
Nuclear Physics, Division of High Energy Physics, of the U.S.\ Department of
Energy under Contract DE-AC03-76SF00098 and by a DOE Outstanding Junior
Investigator award.   M.B.W.~was supported in part by the Department of Energy
under  Grant No.~DE-FG03-92-ER40701.


\end{document}